\documentclass[aps,prd,twocolumn,groupedaddress,showpacs]{revtex4}
\usepackage{graphicx}
\usepackage{dcolumn}
\usepackage{bm}
\usepackage{amsmath}
\usepackage{epstopdf}
\usepackage{amsfonts}
\usepackage{amssymb}%
\usepackage{hyperref} 

\usepackage{natbib}

\providecommand{\U}[1]{\protect\rule{.1in}{.1in}}

\newcommand{\baa}{\begin{align}}
\newcommand{\eaa}{\end{align}}

\newcommand{\be}{\begin{equation}}
\newcommand{\ee}{\end{equation}}
\newcommand{\bea}{\begin{eqnarray}}
\newcommand{\eea}{\end{eqnarray}}

\begin{document}



\title{Greybody factors for a minimally coupled massless scalar field \\
in Einstein-Born-Infeld dilaton spacetime}


\author{Grigoris Panotopoulos}
\affiliation{CENTRA, Instituto Superior T{\'e}cnico, Universidade de Lisboa, Av. Rovisco Pa{\'i}s 1, Lisboa, Portugal}
\email{grigorios.panotopoulos@tecnico.ulisboa.pt}

\author{\'Angel Rinc\'on}
\affiliation{Instituto de F\'{i}sica, Pontificia Universidad Cat\'{o}lica de Chile, \mbox{Avenida Vicu\~na Mackenna 4860, Santiago, Chile.}}
\email{arrincon@uc.cl}

%

\date{\today}

\begin{abstract}
We have analyzed in detail the propagation of a minimally coupled massless scalar field
in the gravitational background of a four-dimensional Einstein-Born-Infeld dilaton
charged black hole. We have obtained analytical expressions for the absorption cross section as well as for the decay rate for the scalar field in the aforementioned spacetime, and we have shown graphically its behavior for different values of the free parameters of the theory.
\end{abstract}

\maketitle


\section{Introduction}
Black holes are objects of paramount importance in gravitational theories. Of particular interest
is Hawking's radiation \cite{hawking}. Since it is a manifestation of a quantum effect
in curved spacetime, Hawking's radiation has always attracted a lot of attention in the community,
despite the fact that up to now it has never been detected in the universe.
The greybody factor, or else the absorption cross section, is a frequency dependent factor that
measures the modification of the original black body radiation, and thus gives us valuable information
about the near horizon structure of black holes \cite{kanti}. Consequently, in the literature exist many
works in which the authors have studied the propagation and the relativistic scattering of different kinds of
fields, and have analyzed the corresponding greybody factors.

Relativistic scattering of waves has been traditionally studied in asymptotically flat spacetimes without a cosmological constant \cite{collection}. However,
due to inflation \cite{guth}, the current cosmic acceleration \cite{riess} and the AdS/CFT correspondence \cite{adscft}, asymptotically non-flat spacetimes
with a positive or negative cosmological constant have also been studied over the years \cite{3D,Fernando:2004ay,coupling,kanti2,Panotopoulos:2016wuu,Ahmed:2016lou}.
In \cite{dilatonBI}, however, the authors have found black hole solutions in three and four dimensions that are neither asymptotically flat nor
asymptotically (anti) de Sitter. In those works the model is described by the Einstein-Born-Infeld dilaton action. Originally the Born-Infeld non-linear
electrodynamics was introduced in the 30's in order
to obtain a finite self-energy of point-like charges \cite{BI}. During the last decades this type of action reappears in the open sector of superstring theories \cite{stringtheory} as it describes the dynamics of D-branes \cite{Dbranes}. Furthermore, in the closed sector of all superstring theories at the massless level the graviton is accompanied by the dilaton that determines the string coupling constant. Since superstring theory is so far the only consistent theory of quantum gravity, it is more than natural to study the greybody factors of black hole solutions obtained in the framework of Einstein-Born-Infeld dilaton models.

In this work we wish to find analytical expressions for the reflection coefficient,
the absorption cross-section and the decay rate for a minimally coupled massless scalar field in a four-dimensional Einstein-Born-Infeld dilaton
spacetime. Our work is organized as follows: After this introduction, we present the model and the wave equation in the next section. In section 3 we
obtain exact solution of the radial equation in terms of hypergeometric functions, and we compute the reflection coefficient as well as the
absorption cross-section and the decay rate in section 4. Finally, we conclude our work in the last section.

\section{Wave equation of a massless scalar field in the Eistein-Born-Infeld dilaton spacetime}
Our starting point is the model considered in the second paper of \cite{dilatonBI} described by the action
\begin{align}
S = \int d^4 x \sqrt{-g} \Bigl[ &R-2(\nabla \phi)^2-V(\phi) \ +
\\
&4 \gamma e^{-2 \kappa \phi} (1-\sqrt{1+Y}) \Bigl]\nonumber
\end{align}
where
\begin{equation}
Y = \frac{F_{\mu \nu} F^{\mu \nu}}{2 \gamma}
\end{equation}
with the scalar field $\phi$ being the dilaton, $V(\phi)$ its potential, $\gamma$ the Born-Infeld parameter,
$\kappa$ the dilaton coupling constant, and $F_{\mu \nu}$ the electromagnetic field strength.
The solution for the dilaton is given by \cite{dilatonBI}
\begin{equation}
\phi(r) = \frac{\kappa}{1+\kappa^2} \ln(br-c)
\end{equation}
where $b,c$ are constants of integration. In the following we set for convenience and without loss of generality $b=1$ and $c=0$.
On the other hand the line element for the metric is given by \cite{dilatonBI}
\begin{equation}
ds^2=-h(r) dt^2 + h(r)^{-1} dr^2 + e^{2 \kappa \phi} d \Omega^2
\end{equation}
The dilaton potential is taken to be either $V(\phi)=0$
or a Liouville type potential $V(\phi)=2 \Lambda e^{-2 \kappa \phi}$. Since the model is string inspired, in the following
we shall consider the string coupling case $\kappa=1$, in which $h(r)$ is given either by \cite{dilatonBI}
\begin{equation}
h(r) = 2r \left(1-2H-\frac{r_0}{2r} \right)
\end{equation}
if $V(\phi)=0$, or by \cite{dilatonBI}
\begin{equation}
h(r) = 2r \left(1-2H-\Lambda-\frac{r_0}{2r} \right)
\end{equation}
if $V(\phi)=2 \Lambda e^{-2 \phi}$. Therefore in both cases the function $h(r)$ turns out to be linear in $r$ irrespectively of
the dilaton potential, namely $h(r)=r/L - r_0$ where $r_0,L$ are
constants. The constant $r_0$ is related to the mass of the black hole \cite{dilatonBI}, $r_0=4M$, while the length scale $L$
is given by
\begin{equation}
L^{-1} = 2 (1-\Lambda-2 H)
\end{equation}
where the constant $H$ is given by \cite{dilatonBI}
\begin{equation}
H = -\gamma + \sqrt{\gamma (Q^2+\gamma)}
\end{equation}
and the charge $Q$ of the black hole is given by \cite{dilatonBI}
\begin{equation}
Q^2 = \frac{1+\sqrt{1+16 \gamma^2}}{8 \gamma}
\end{equation}
There is a single event horizon $r_H=L r_0$, and therefore the line element takes the form
\begin{equation}
ds^2=-h(r) dt^2 + h(r)^{-1} dr^2 + r d \Omega^2
\end{equation}
where $h(r)=(r-r_H)/L$. Therefore the metric is characterized by two length scales $L,r_H$ which are given functions of the three
free parameters of the model, namely the black hole mass $M$, the Born-Infeld parameter $\gamma$, and the mass scale $\Lambda$ in dilaton's potential.
Note that $r_H$ depends on all three free parameters, while $L$ does not depend on the mass of the black hole.

Now we consider a minimally coupled massless scalar field $\Psi$ in the above background. The equation of motion
is the standard Klein-Gordon equation
\begin{equation}
\frac{1}{\sqrt{-g}} \partial_\mu (\sqrt{-g} g^{\mu \nu} \partial_\nu \Psi) = 0
\end{equation}
and using the ansatz $\Psi(t,r,\theta, \phi)=e^{-i \omega t} R(r) Y_l^m(\theta, \phi)$,
where $Y_l^m$ are the usual spherical harmonics, we obtain the radial equation
\begin{equation} \label{radial}
R'' + \left(\frac{h'}{h}+\frac{1}{r}\right) R' + \left(\frac{\omega^2}{h^2}-\frac{l (l+1)}{r h}\right) R = 0
\end{equation}
To see the potential that the scalar field feels we define new variables as follows
\begin{eqnarray}
R & = & \frac{\psi}{\sqrt{r}} \\
x & = & \int \frac{dr}{h(r)}=L \ln\Bigl(\frac{r-r_H}{d}\Bigl)
\end{eqnarray}
where $x$ is the so called tortoise coordinate and $d$ is a constant of integration which will be taken as unity. We recast the equation for the radial part into a Schr{\"o}dinger-like equation of the form
\begin{equation}
\frac{d^2 \psi}{dx^2} + (\omega^2 - V(x)) \psi = 0
\end{equation}
Therefore we obtain for the effective potential barrier the expression
\begin{equation}
V(r) = h(r) \: \left( \frac{l (l+1)}{r}+\frac{h'(r)}{2 r}-\frac{h(r)}{4 r^2} \right)
\end{equation}
which can be simplified to be
\begin{equation}
V(r) = V_0 - \frac{r_H l (l+1)}{L r} - \frac{r_H^2}{4 L^2 r^2}
\end{equation}
where the constant term is given by $V_0=(Ll(l+1)+1/4)/L^2$. The effective potential barrier as a function of the radial distance can be
seen in Fig. 1 and 2 below for $l=1$ and $l=0$ respectively, and three different cases, namely $\gamma=1$, $\Lambda=0$, and $M=1.1, 1.75, 2.25$.

Since at the horizon the effective potential vanishes, the general solution for the function $\psi$ close to the horizon (where $\omega^2 \gg V(x)$)
is given by
\begin{equation}
\psi(x) = C_{+} e^{i \omega x} + C_{-} e^{-i \omega x}
\end{equation}
while requiring purely ingoing solution we set $C_{-}=0$, and thus the solution becomes
\begin{equation}\label{ingoing}
\psi(x) = C_{+} e^{i \omega x}
\end{equation}
On the other hand, it is easy to check that at large $r$ (or at large $x$, since when $r \gg r_H$, $r \simeq e^{x/L}$) the potential tends to
the constant $V_0$, and therefore
defining $\Omega \equiv \sqrt{\omega^2-V_0}$ the solution
for $\psi$ is given by
\begin{align}
\psi(x) &= A_{+} e^{i \Omega x} + A_{-} e^{-i \Omega x}.
\end{align}
Therefore the far-field solution expressed in the tortoise coordinate $x$ takes the form of ingoing and outgoing plane waves provided that
$\omega^2 > V_0$.

\begin{figure}[ht!]
\centering
\includegraphics[width=\linewidth]{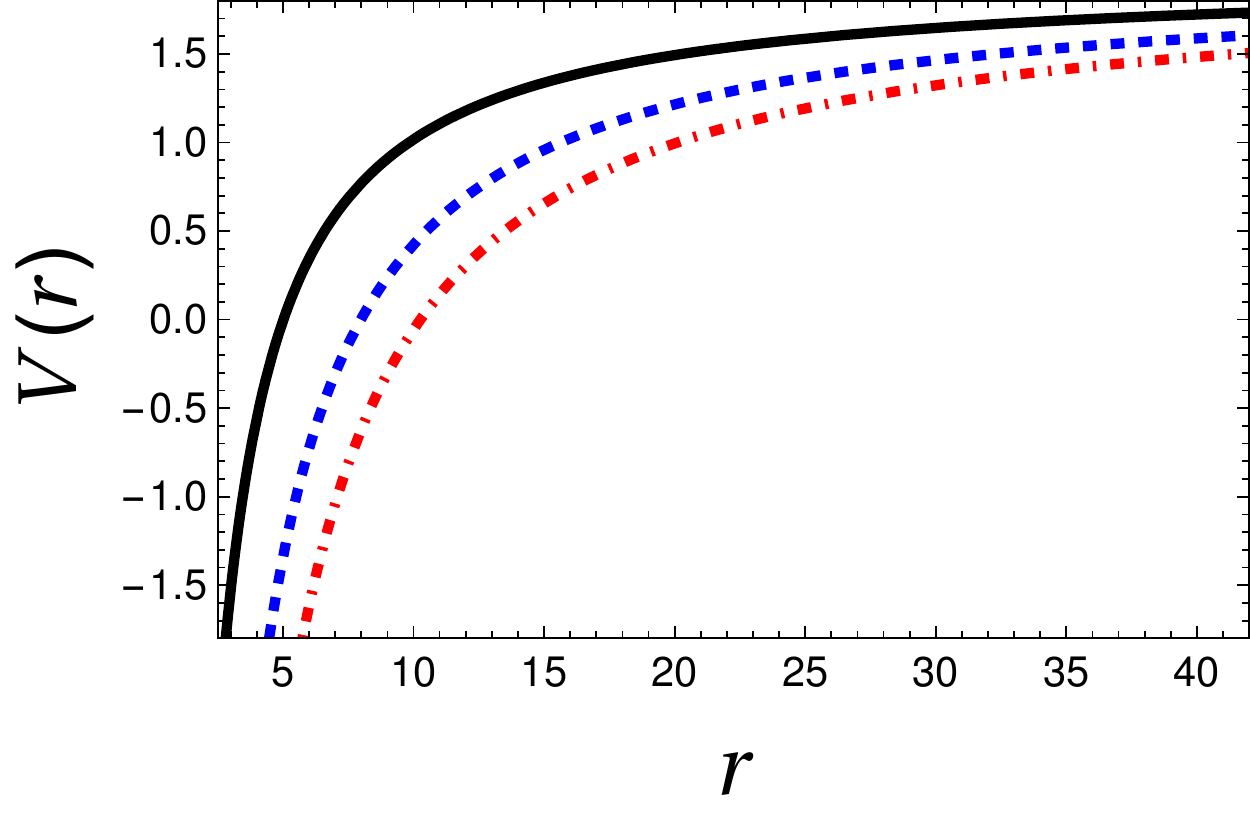}
\caption{Effective potential versus $r$ for $l = 1, \gamma=1, \Lambda=0$ and for $M=1.1$ (solid black line), $M=1.75$ (dashed blue line) and $M=2.25$ (dot-dashed red line).}
\end{figure}

\begin{figure}[ht!]
\centering
\includegraphics[width=\linewidth]{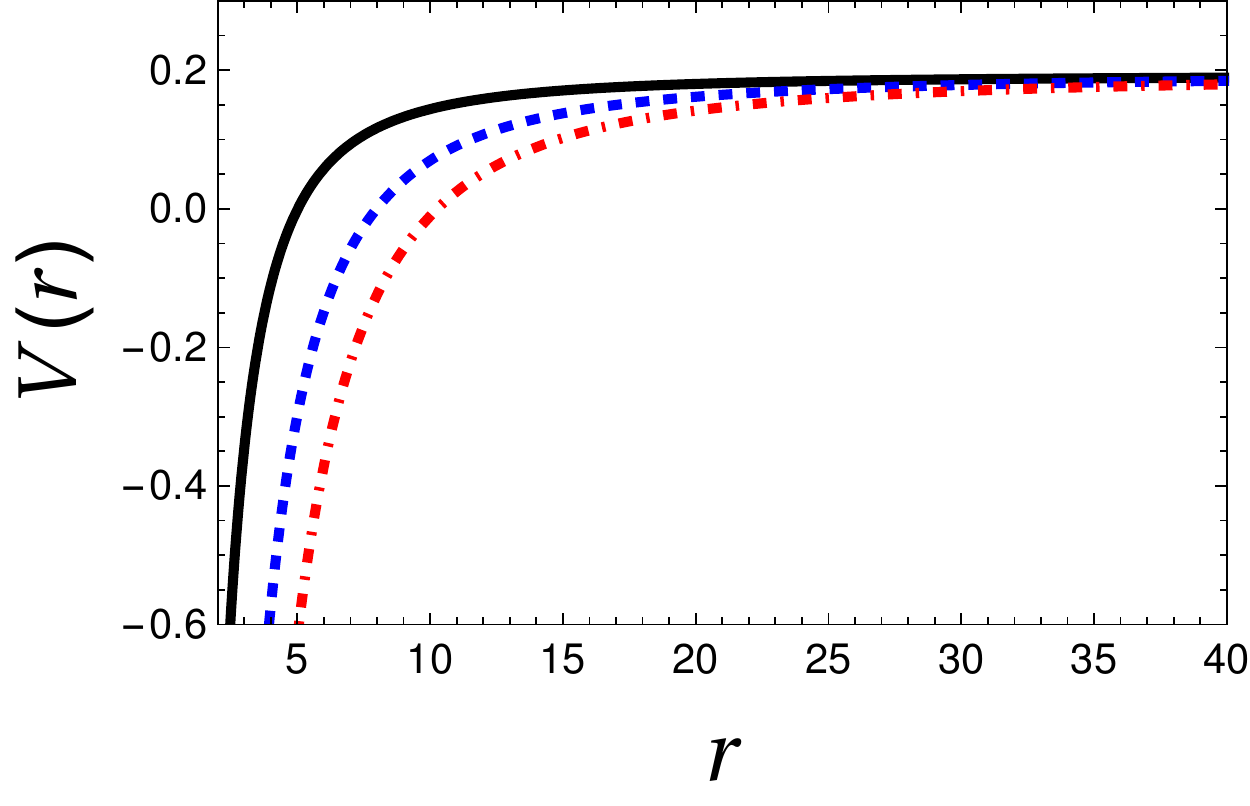}
\caption{Same as in Fig. 1, but for $l = 0$.}
\end{figure}

\section{Solution of the radial differential equation}

\subsection{Solution in the far-field region}

In the far-field region $r \gg r_H$ $h(r) \sim r/L$ and the radial equation becomes
\begin{equation}
R''(r) + \frac{2}{r} R'(r) + \frac{q}{r^2} R(r)= 0
\end{equation}
where $q=(\omega L)^2-L l (l+1)$
or
\begin{equation}
r^2 R''(r) + 2 r R'(r) + q R(r) = 0
\end{equation}
which is Euler's equation. We seek solutions of the form $R(r) \sim r^{\rho}$ and the power
$\rho$ satisfies the algebraic equation
\begin{equation}
\rho^2+\rho+q = 0
\end{equation}
The determinant is found to be $\Delta = -4L^2 (\omega^2-V_0)$, and the algebraic equation above admits two roots given by
\begin{equation}\label{rho}
\rho_{\pm} = -\frac{1}{2} \pm \sqrt{\frac{1}{4}-q}
\end{equation}
and they are real when $1/4 \geq q$, while when $1/4 < q$ the roots are complex.
Therefore the solution in the far-field region reads
\begin{equation}
R_{FF}=D_1 \left( \frac{r}{r_H} \right)^{\rho_-} + D_2 \left( \frac{r}{r_H} \right)^{\rho_+}
\end{equation}
where $D_1, D_2$ are two arbitrary coefficients. In terms of the tortoise coordinate $x$, the radial part takes the form of ingoing and outgoing
plane waves only when the two roots above are complex as follows
\begin{align}
R_{FF} &= \frac{1}{\sqrt{r}} \left ( \frac{D_1}{r_H^{\rho_{-}}}\ e^{-i \Im{(\rho_{+})}\frac{x}{L}}
+
\frac{D_2}{r_H^{\rho_{+}}}\ e^{i \Im{(\rho_{+})}\frac{x}{L}} \right )
\end{align}
with $\Im{(\rho_{+})}$ being the imaginary part of $\rho_{+}$ defined in Eq. (\ref{rho}). It can also be seen from the solution
of the Schr{\"o}dinger-like equation at large $x$, as we already mentioned in the end of Section II. From the above it is clear that $D_1,D_2$ represent
the ingoing and outgoing waves respectively. Therefore the reflection coefficient is defined to be $\mathcal{R} = |D_1/D_2|^2$, and
in the following we shall consider the case where $q > 1/4$ or $\omega^2 > V_0$. In this case the roots are given by
\begin{equation}
\rho_{\pm} = -\frac{1}{2} \pm i \sqrt{-\frac{1}{4}+q}
\end{equation}
Note that in the far-field solution for the radial part there is also a decaying amplitude $1/\sqrt{r}$, but since it is present in both
terms of the solution, it drops from the final expression upon taking the ratio to compute the reflection coefficient.

\subsection{Exact solution in terms of hypergeometric functions}

Next we find an exact solution of the radial equation (\ref{radial}) in terms of hypergeometric functions introducing $z=1-r_H/r$. The new
equation for $z$ reads
\begin{equation}
z (1-z) R_{zz} + (1-z) R_z + \left( \frac{A}{z} + \frac{B}{-1+z} \right) R = 0
\end{equation}
where $A=(\omega L)^2, B=-(\omega L)^2+L l (l+1)$. To get rid of the poles we set
\begin{equation}
R = z^\alpha (1-z)^\beta F
\end{equation}
where now $F$ satisfies the following differential equation
\begin{multline}
z (1-z) F_{zz} + [1+2 \alpha - (1+2 \alpha+2 \beta) z] F_z
\\
+ \left( \frac{\bar{A}}{z} + \frac{\bar{B}}{-1+z} - C \right) F = 0
\end{multline}
and the new constants are given by
\begin{eqnarray}
\bar{A} & = & A + \alpha^2 \\
\bar{B} & = & B + \beta - \beta^2 \\
C & = & (\alpha+\beta)^2
\end{eqnarray}
Demanding that $\bar{A} = 0 = \bar{B}$ we obtain the
Gauss' hypergeometric equation
\begin{equation}
z (1-z) F_{zz} + [c-(1+a+b) z] F_z - ab F = 0
\end{equation}
and we determine the parameters $\alpha, \beta$ as follows
\begin{eqnarray}
\alpha & = & i \omega L \\
\beta & = & \frac{1}{2} + i \sqrt{(\omega L)^2 - L l (l+1) -\frac{1}{4}} \label{beta}
\end{eqnarray}
Finally the three parameters of Gauss' equation are given by
\begin{eqnarray}
c & = & 1+2 \alpha \\
a & = &  \alpha + \beta \\
b & = & \alpha + \beta
\end{eqnarray}
Note that the parameters $a,b,c$ satisfy the condition $c-a-b=1-2 \beta$.
Therefore the solution for the radial part is given by
\begin{equation}
R(z) = D z^\alpha (1-z)^\beta F(a,b;c;z)
\end{equation}
where $D$ is an arbitrary coefficient, and the hypergeometric function can be expanded in a Taylor series
as follows
\begin{equation}
F(a,b;c;z) = 1 + \frac{a b}{c} \:z + ...
\end{equation}
Note that the above solution for the choice of $\alpha=i \omega L$ reproduces the purely ingoing solution at the horizon (\ref{ingoing}), as it can be
seen from the fact that close to the horizon ($z \rightarrow 0$), the radial part becomes $R(z) \simeq D z^\alpha$, and
the parameter z can be written approximately $z \simeq (r-r_H)/r_H = e^{x/L}/r_H$.

\subsection{Matching of the solutions}

In order to match with the far field solution obtained earlier (where now $z \rightarrow 1$) we use the transformation \cite{handbook}
\begin{equation}
\begin{split}
F(a,b;c;z) = \ &\frac{\Gamma(c) \Gamma(c-a-b)}{\Gamma(c-a) \Gamma(c-b)}
\ \times
\\
&F(a,b;a+b-c+1;1-z) \ +
\\
 (1-z)^{c-a-b} &\frac{\Gamma(c) \Gamma(a+b-c)}{\Gamma(a) \Gamma(b)}
\ \times
\\
&F(c-a,c-b;c-a-b+1;1-z)
\end{split}
\end{equation}
and therefore the radial part as $z \rightarrow 1$ reads
\begin{equation}
\begin{split}
R(z \rightarrow 1)=D (1-z)^\beta \frac{\Gamma(1+2 \alpha) \Gamma(1-2 \beta)}{\Gamma(1+\alpha-\beta) \Gamma(1+\alpha-\beta)} \\
+D (1-z)^{1-\beta} \frac{\Gamma(1+2 \alpha) \Gamma(-1+2 \beta)}{\Gamma(\alpha+\beta) \Gamma(\alpha+\beta)}
\end{split}
\end{equation}
Note that $-\beta = \rho_{-}$ and $\beta-1=\rho_{+}$, and since $z=1-(r_H/r)$ the radial part $R(r)$ for $r \gg r_H$ can be written down as follows
\begin{equation}
\begin{split}
R(r) = D \frac{\Gamma(1+2 \alpha) \Gamma(1-2 \beta)}{\Gamma(1+\alpha-\beta) \Gamma(1+\alpha-\beta)}
\left(\frac{r}{r_H}\right)^{\rho_{-}} \\
+D \frac{\Gamma(1+2 \alpha) \Gamma(-1+2 \beta)}{\Gamma(\alpha+\beta) \Gamma(\alpha+\beta)} \left(\frac{r}{r_H}\right)^{\rho_{+}}
\end{split}
\end{equation}
Finally upon comparison we express $D_1, D_2$ in terms of $D$ as follows
\begin{eqnarray}
D_1 & = & D \: \frac{\Gamma(1+2 \alpha) \Gamma(1-2 \beta)}{\Gamma(1+\alpha-\beta) \Gamma(1+\alpha-\beta)} \\
D_2 & = & D \: \frac{\Gamma(1+2 \alpha) \Gamma(-1+2 \beta)}{\Gamma(\alpha+\beta) \Gamma(\alpha+\beta)}
\end{eqnarray}

\section{The absorption cross-section and discussion of the results}

The reflection coefficient is given by $\mathcal{R} = |D_1/D_2|^2$, and according to the previous results is computed to be
\begin{align}
\mathcal{R} &=
\left| \frac{\Gamma(\alpha+\beta)^2 \Gamma(1-2 \beta)}{\Gamma(1+\alpha-\beta)^2 \Gamma(-1+2 \beta)} \right|^2
\end{align}
and using the following identities for the $\Gamma$ function \cite{Fernando:2004ay}
\begin{eqnarray}
\Bigl|\Gamma(i y)\Bigl|^2 & = & \frac{\pi}{y \sinh(\pi y)}  \\
\left|\Gamma\left(\frac{1}{2}+i y\right)\right|^2 & = & \frac{\pi}{\cosh(\pi y)}
\end{eqnarray}
we obtain the final expression
\begin{equation}
\mathcal{R} = \frac{\cosh ^2\left[\pi  \left[L \omega -\sqrt{-l (l+1) L+L^2 \omega^2-\frac{1}{4}}\right]\right]}{
 \cosh^2\left[\pi  \left[L \omega + \sqrt{-l (l+1) L+L^2 \omega^2-\frac{1}{4}}\right]\right]}
\end{equation}
The reflection coefficient depends on $L$ only, and not on $r_H$. Therefore it does not depend on the mass of the black hole. As a function of the frequency it can be seen in Fig. 3 and 4 below for $l=0$. It always starts
at 1 and monotonically decreases to zero very fast. In Fig. 3 we have set $\Lambda=0$ and we have considered three different values for $\gamma=0.5, 1, 2$, while in Fig. 4 we have set $\gamma=1$ and we have considered three different values for $\Lambda=0.001, 0.01, 0.1$. The curves move to the right as we decrease $L$ (decrease $\Lambda$ or increase $\gamma$) due to the inequality found before
$\omega > \sqrt{4 l^2 L+4 l L+1}/2L$.

\begin{figure}[ht!]
\centering
\includegraphics[width=\linewidth]{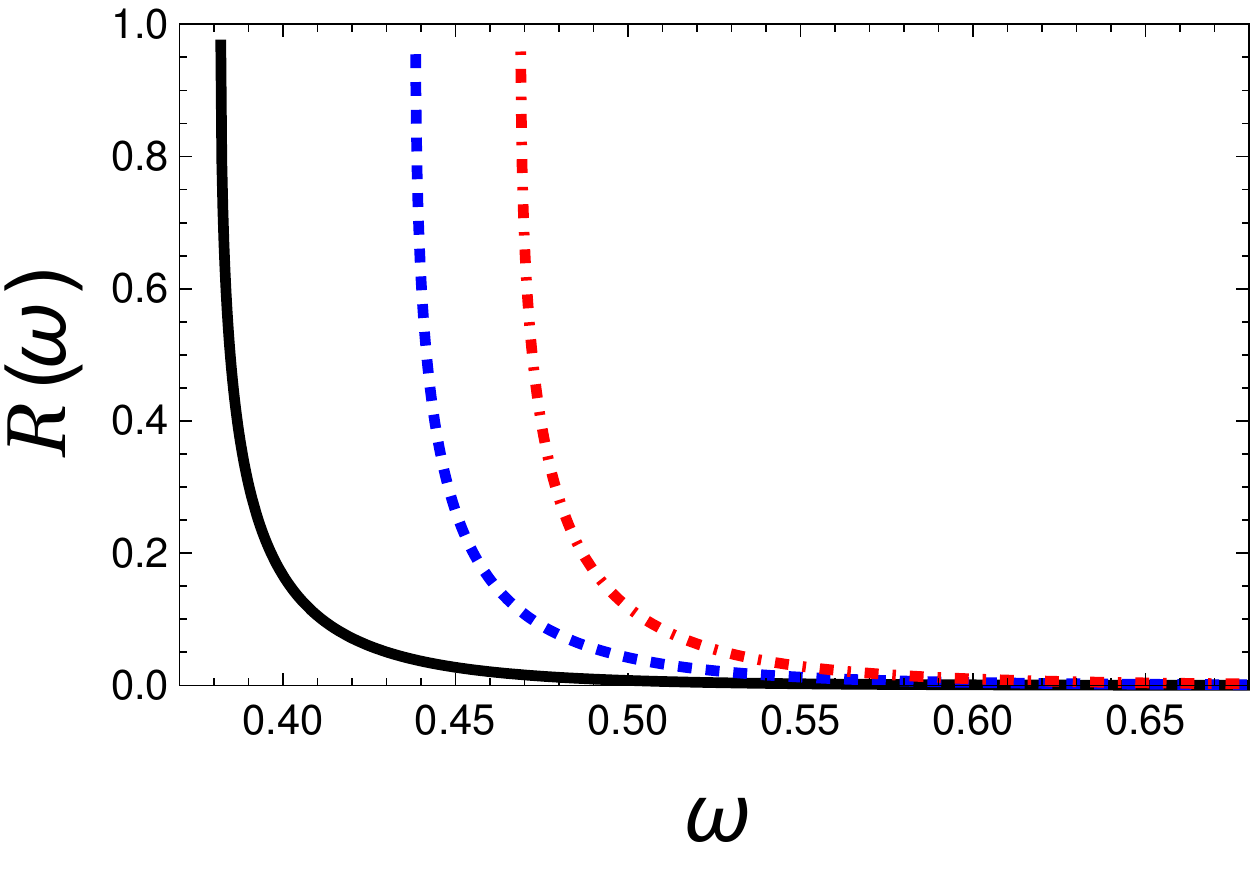}
\caption{
Reflection coefficient versus $\omega$ for $l = 0$, $\Lambda=0$ and from left to right $\gamma = 0.5$, $\gamma = 1$ and $\gamma = 2$.
}\label{R2}
\end{figure}

\begin{figure}[ht!]
\centering
\includegraphics[width=\linewidth]{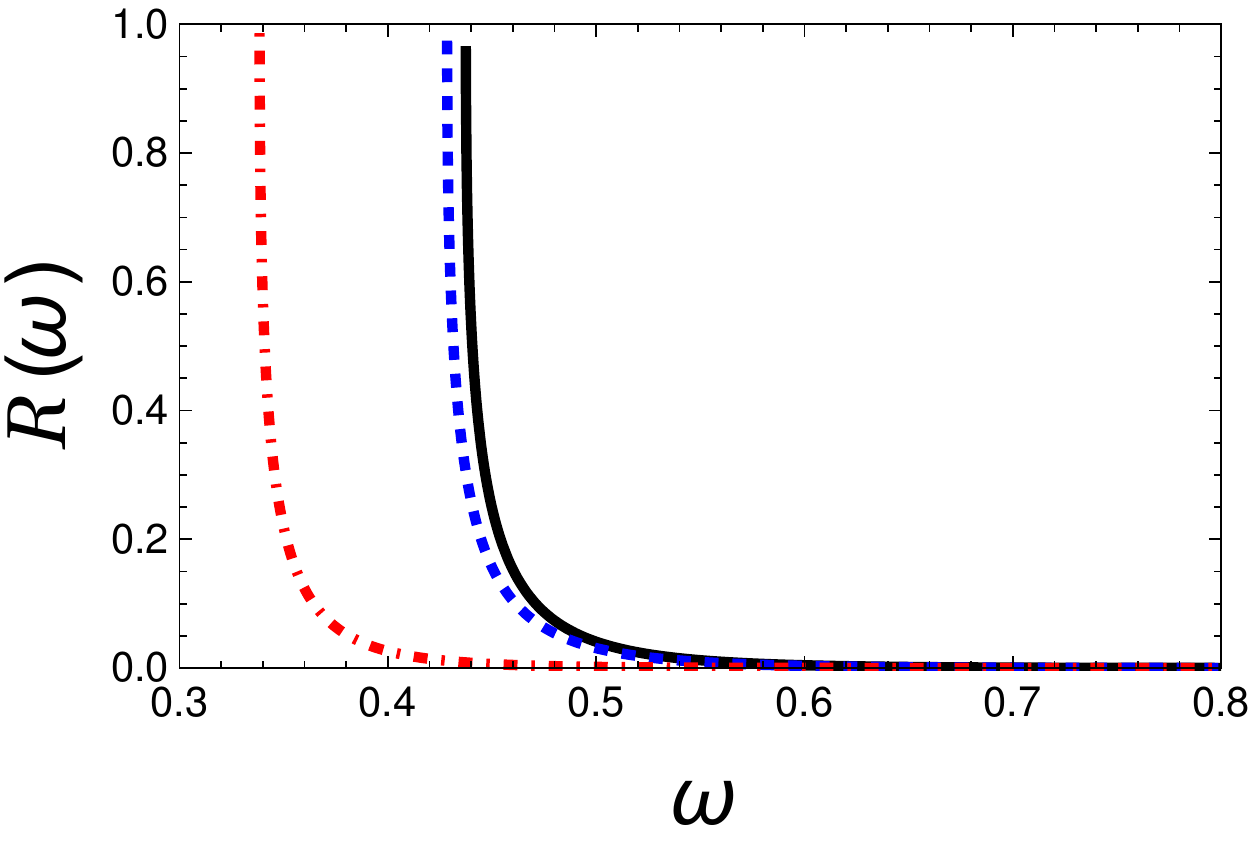}
\caption{
Reflection coefficient versus $\omega$ for $l = 0$, $\gamma = 1$, and from left to right $\Lambda = 0.1$, $\Lambda = 0.01$ and $\Lambda=0.001$.
}\label{R2l}
\end{figure}

The absorption cross section is given by the optical theorem \cite{optical,kanti}
\begin{equation}
\sigma_{abs} = \Lambda_{l} \frac{(1-\mathcal{R})}{\omega^2}
\end{equation}
where $\Lambda_{l} = (2l+1) \pi$ and therefore the full expression is given by
\begin{align}
\sigma_{abs} &= \frac{\Lambda_{l}}{\omega^2} \left[ 1 - \frac{\cosh ^2\left[\pi  \left[L \omega - \Im(\beta)\right]\right]}{
 \cosh^2\left[\pi  \left[L \omega + \Im(\beta)\right]\right]}  \right]
\end{align}
where $\Im(\beta) = \Im{(\rho_{+})}$ denotes the imaginary part of $\beta$ defined at Eq. (\ref{beta}).
The greybody factor as a function of $\omega$ is shown in Fig. 5 and 6 below for $l=0$. We have considered
the same values for $\gamma$ and $\Lambda$ as in Fig. 3 and 4 respectively.
First the absorption cross section increases with $\omega$ until it reaches a maximum, and then tends to zero monotonically but not
as fast as the reflection coefficient. Due to the same inequality, as we decrease $L$ the curves move to the right, while at the same
time the maximum value gets lower.

Since the flux spectrum emitted by the black hole is given by \cite{coupling}
\begin{equation}
\frac{dN(\omega)}{dt} = \sum_{l} \frac{\sigma_{l}(\omega)}{e^{\omega/T_H} - 1} \frac{d^3 k}{(2 \pi)^3}
\end{equation}
the decay rate is defined to be \cite{Fernando:2004ay}
\begin{equation}
\Gamma_{decay} = \frac{\sigma_{abs}}{e^{\omega/T_H}-1}
\end{equation}
where the Hawking temperature is computed to be $T_H=1/(4 \pi L)$ \cite{dilatonBI}.
Thus, the full expression is given by
\begin{align}
\Gamma_{decay} &= \frac{\Lambda_{l}}{\omega^2}
\left[
\frac{e^{2 \pi  \sqrt{-4 L \left(l^2+l-L \omega^2\right)-1}}-1}{\left[e^{\pi \sqrt{-4 L \left(l^2+l-L \omega^2\right)-1} + 2 \pi L \omega }+1\right]^2}
\right]
\end{align}
and as a function of the frequency it can be seen in Fig. 7 and 8 below for $l=0$. We have considered the same values for $\gamma$ and $\Lambda$ as in Fig. 3 and 4 respectively. The decay rate reaches a maximum and then quickly decays to zero. As we decrease $\Lambda$ or increase $\gamma$ the curves move to the right and at the same time the maximum value decreases too, precisely as in the greybody factor case.

\begin{figure}[ht!]
\centering
\includegraphics[width=\linewidth]{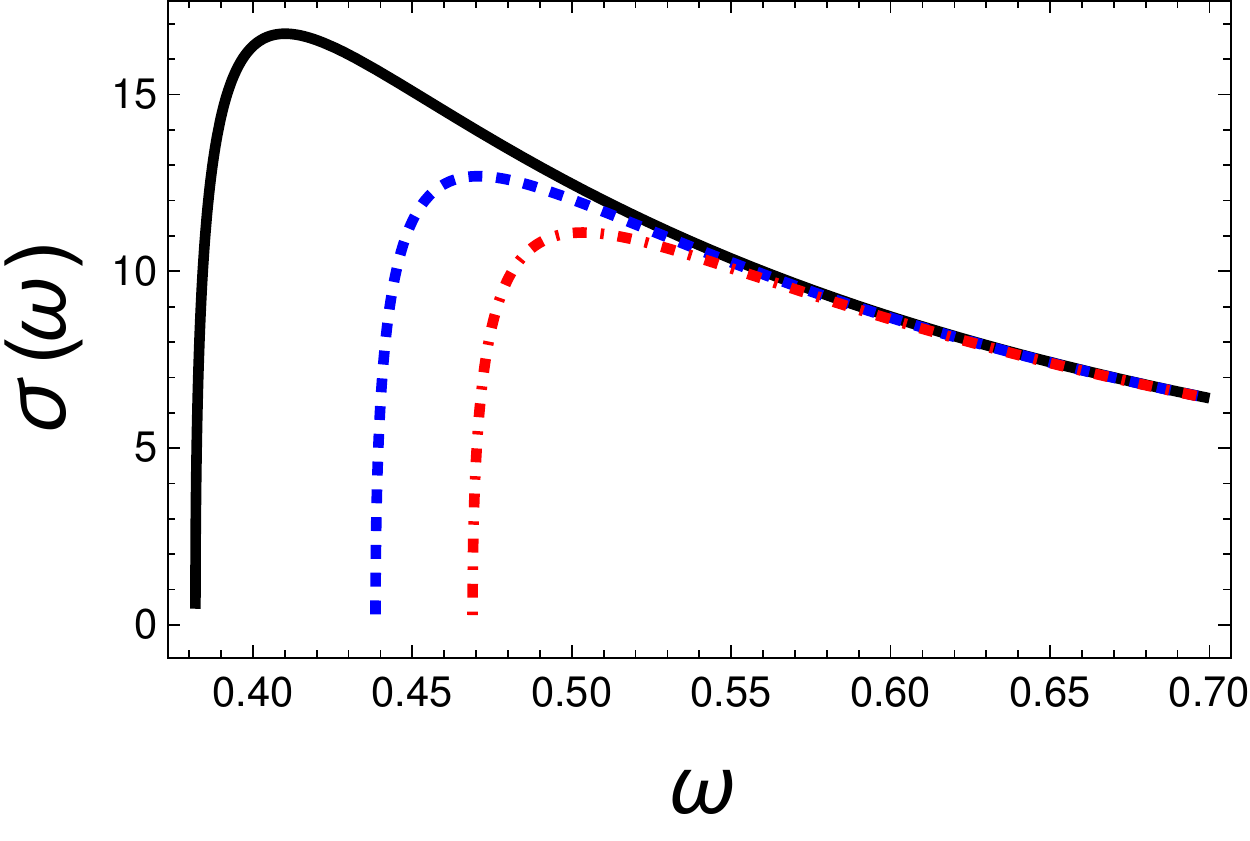}
\caption{
Absorption cross section as a function of $\omega$ for $l = 0$ and values for $\gamma, \Lambda$ as in Fig. 3.
}\label{sigma}
\end{figure}

\begin{figure}[ht!]
\centering
\includegraphics[width=\linewidth]{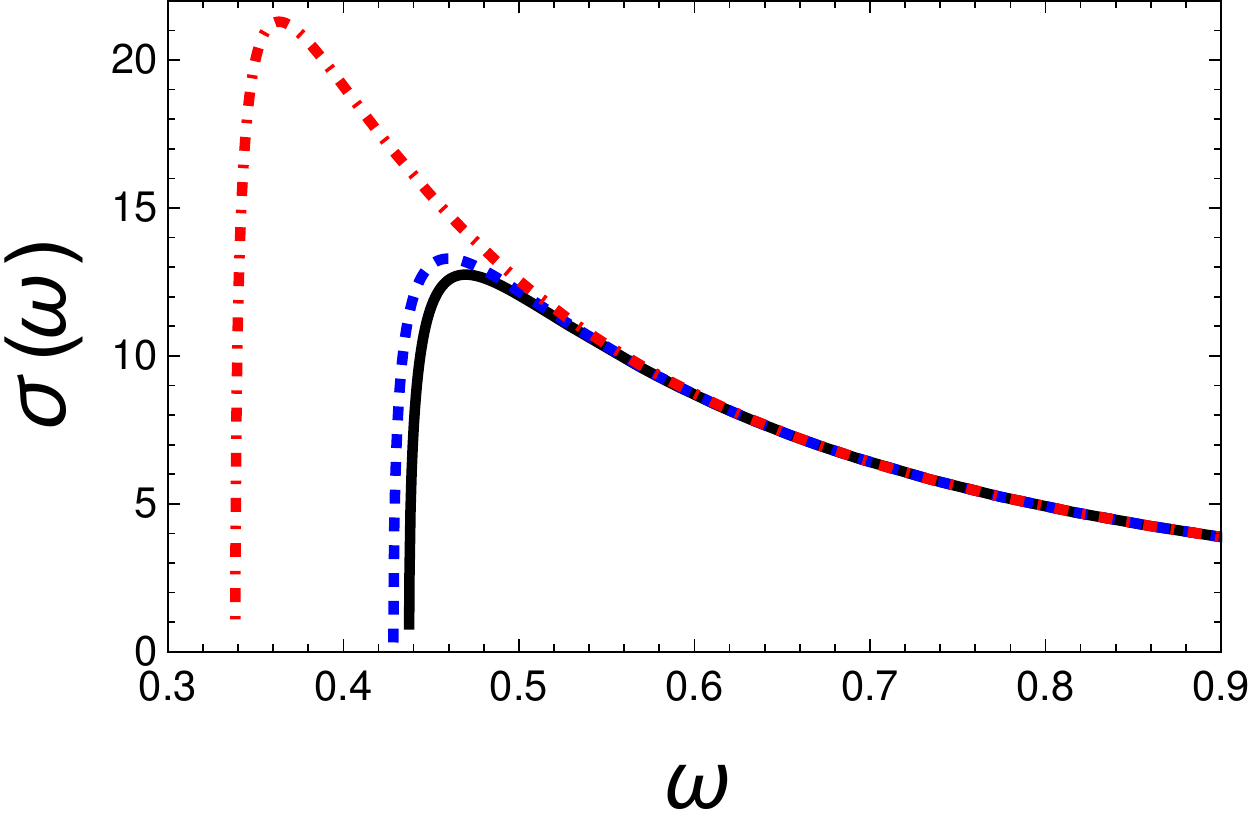}
\caption{
Absorption cross section as a function of $\omega$ for $l = 0$ and values for $\gamma, \Lambda$ as in Fig. 4.
}\label{sigmal}
\end{figure}

\begin{figure}[ht!]
\centering
\includegraphics[width=\linewidth]{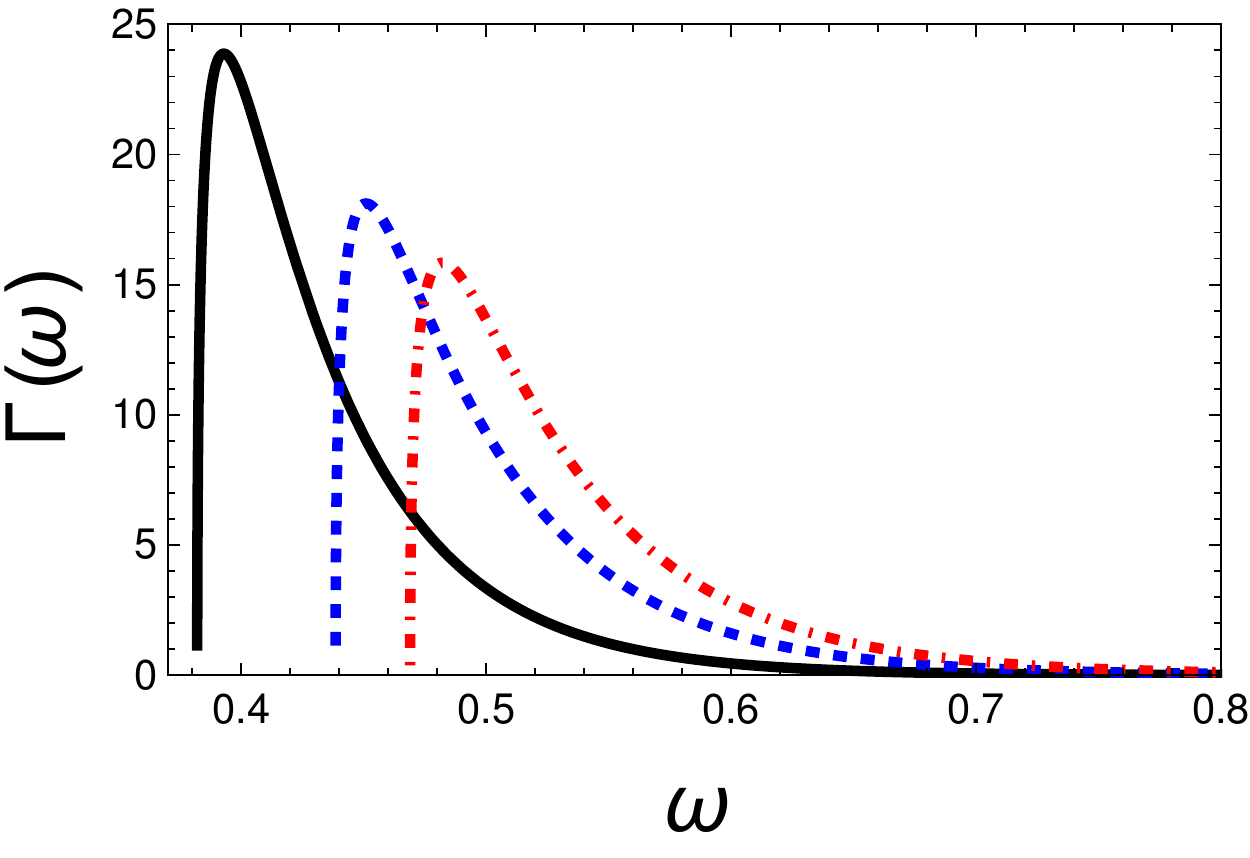}
\caption{
Decay rate as a function of $\omega$ for $l = 0$ and values for $\gamma, \Lambda$ as in Fig. 3. Note that the vertical axis is scaled $1 : 10^{-3}$.
}\label{Gamma}
\end{figure}

\vspace{10cm}

\begin{figure}[ht!]
\centering
\includegraphics[width=\linewidth]{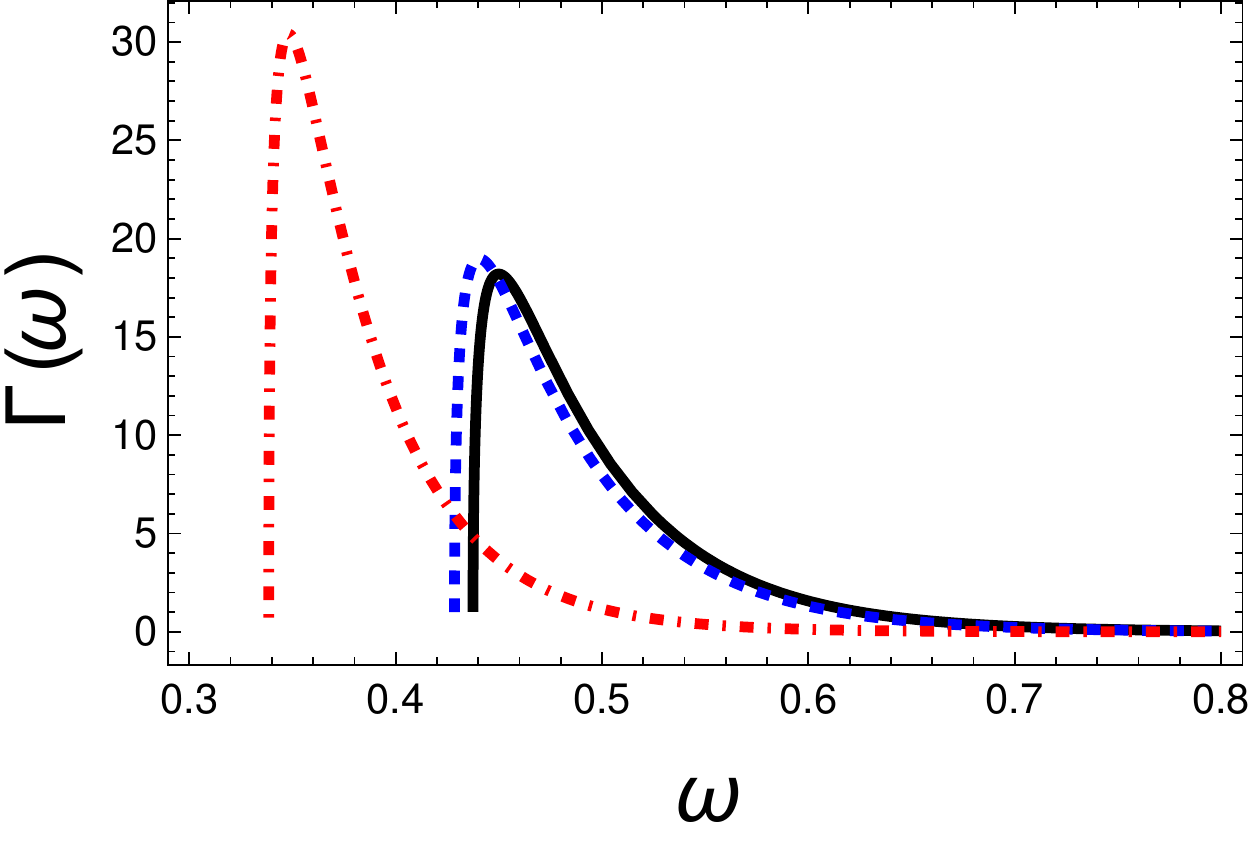}
\caption{
Decay rate as a function of $\omega$ for $l = 0$ and values for $\gamma, \Lambda$ as in Fig. 4.  Note that the vertical axis is scaled $1 : 10^{-3}$.
}\label{Gammal}
\end{figure}

\section{Conclusions}

To summarize, in this article we have analyzed the greybody factors for a minimally coupled massless
scalar scalar field in a four-dimensional Einstein-Born-Infeld dilaton charged black hole background.
Since the model is string inspired we have considered the string coupling case $\kappa=1$. We have found exact solution of
the radial equation in terms of the hypergeometric functions, and we have obtained
analytical expressions for the effective barrier potential, the reflection coefficient,
the absorption cross section as well as the decay rate.
We have shown in figures how the above quantities behave for different values of the free parameters of the theory. Our results are
qualitatively similar to those for a three-dimensional Einstein-Maxwell dilaton black hole.


\begin{acknowledgments}

We wish to thank the anonymous reviewers for valuable comments and suggestions. The author G.P. acknowledges the support
from "Funda{\c c}{\~a}o para a Ci{\^e}ncia e Tecnologia". The work of A.R. was supported by the CONICYT-PCHA/Doctorado Nacional/2015-21151658.
\end{acknowledgments}
%
%
%


%
%

\end{document}